\documentstyle[editedvolume]{crckapbk.sty}

\newcommand{\kms}{$\,{\rm km\,s^{\scriptscriptstyle -1}}$}
\newcommand{\gtsim}{\ {\raise-0.5ex\hbox{$\buildrel>\over\sim$}}\ }
\newcommand{\ltsim}{\ {\raise-0.5ex\hbox{$\buildrel<\over\sim$}}\ }
\def\Msun{\hbox{$\thinspace M_{\odot}$}}

\def\deg   {$^\circ$}

\def\etal{{\rm et~al. }}
\def\ltsima{$\; \buildrel < \over \sim \;$}
\def\simlt{\lower.5ex\hbox{\ltsima}}
\def\gtsima{$\; \buildrel > \over \sim \;$}
\def\simgt{\lower.5ex\hbox{\gtsima}}

\begin{opening}

\title{High-Velocity Clouds Related to the Magellanic System}

\author{M. E. Putman}

\institute{Center for Astrophysics and Space Astronomy, University of 
Colorado, Boulder CO, 80309-0389, USA.}

\end{opening}

\runningtitle{Magellanic HVCs}

\begin{document}

\section{Introduction}

  The results of the interaction between the Milky Way and the Magellanic Clouds are 
revealed through several high velocity 
complexes which are connected to the Clouds.  The exact
mechanism of their formation is under some debate, but they remain 
the only group of high-velocity clouds (HVCs) for which we have an origin and roughly a distance. 
Given that, the Magellanic HVCs can be used as a calibrator for 
other HVCs, while also providing an opportunity to
closely investigate the remnants of an interacting system.   These HVCs may hold 
the key to the star formation history,
kinematic structure, and present Hubble type of the Magellanic Clouds, and their 
proximity to the Milky Way allows us to estimate key Galactic parameters.  


The HVCs related to the Magellanic System can be classified into three major complexes:  the Magellanic Bridge, an HI connection between the Clouds;
the Magellanic Stream, which trails the clouds and is one of the largest HI features 
in the sky outside of our Galaxy; 
and the Leading Arm, a more diffuse HI filament which leads
the Clouds.
In terms of HVCs, these features have been studied rather extensively.
In this review, I will first describe the observational results
for each complex and subsequently discuss their origin and 
relationship to the overall HVC population.  All of the HI data
presented are from the HI Parkes All-Sky Survey (HIPASS) (Barnes et al. 2001;
Putman et al. 2001).

\section{The Magellanic Bridge}

\subsection{HI structure and kinematics}

The Magellanic Bridge is a continuous filament of HI which stretches from the
body of Small Magellanic Cloud (SMC) to an extended arm of the Large 
Magellanic Cloud (LMC) (see Figure 1).  The Bridge merges
almost seamlessly with the SMC, but the boundary between the Bridge
and SMC is usually defined at
$(\ell,b)=(295^\circ,-41.5^\circ)$ and $v_{LSR}=125$ \kms, where the loop
which extends from the SMC at approximately $(\ell,b)=(297^\circ,44^\circ)$
rejoins the tail of the SMC (also known as Shapley's wing).  This is
also the boundary that was originally chosen based on stellar associations (Westerlund \& 
Glaspey 1971); however,
with the increasing numbers of stellar associations found in the Bridge, this boundary is also
somewhat ambiguous (see $\S$2.2).  The Bridge emerges from the SMC's tail at the
high column densities of $10^{21}$ cm$^{-2}$ and remains clumpy, but gradually decreases 
in column density to $10^{20}$ cm$^{-2}$ at $(\ell,b)=(287^\circ,-35.5^\circ)$. 
At this latter position, the Bridge joins with what appears to be an extended
spiral arm of the LMC (Kim~\etal 1998; Putman~\etal 1999).
South of the Bridge in Galactic coordinates, especially on the SMC-side, the chaotic beginnings of the
Magellanic Stream are present.
In general, the Bridge is a more orderly feature than the Stream, possibly 
representing the Bridge's shorter history or a more stable environment.  
The Bridge has an HI mass of approximately $5.5 \times 10^7$ \Msun, but this value is
 highly dependent on whether extensions into the SMC, LMC and Stream are included.  


The Bridge has a regular velocity gradient along its main filament,
gradually increasing in velocity and decreasing in spatial width as it approaches the LMC.  The
final dense pockets of emission do not disappear until 350 \kms~at $(\ell,b)=(283^\circ,-42^\circ)$.
McGee \& Newton (1986) report on line profiles which contain up to 5 components
throughout the Bridge (with a velocity resolution of 4.1 \kms).  They report systematic
profile variations in the central Bridge region, but sporadic differences in the regions
of the Bridge which extend into the Magellanic Stream, possibly indicating a more turbulent
environment.  The presentation of high resolution data has begun
with the Australia Telescope Compact Array (ATCA) observations of
Stanimirovic~\etal (1997).  Detailed kinematic information will also soon
be available with Parkes multibeam narrow-band observations.
HI absorption studies find that the cool atomic phase gas exists in the Bridge,
indicating that the pressure in this region is surprisingly high and that
stars may have formed from the Bridge material directly, rather than being drawn out
from the SMC (Kobulnicky \& Dickey 1999).  Sensitive CO studies of the Bridge
would be an interesting future pursuit.

\subsection{Stars!}

The Magellanic Bridge is the only HVC which has stars associated with it, and in this
respect it may be inaccurate to call it an HVC\footnote{Also, although
the Bridge is an HVC in the Galactic reference frame, it is not technically an HVC in the 
Magellanic reference frame, unlike the Stream and Leading Arm.}. 
The stars are very scarce and the gas to
star ratio remains extremely high, so it is conceivable that future stellar searches
may find stars associated with other HVCs.  Early stellar searches in the SMC tail included the
discovery of a number of B-type giants and dwarfs (e.g. 
Sanduleak 1969).  Searches for blue stars then continued throughout the Bridge 
(e.g. Irwin~\etal 1990), and were identified from $(\ell,b)=(296^\circ,-41^\circ)$
to at least $(\ell,b)=(287^\circ,-36^\circ)$.
Demers \& Battinelli (1998) find that
the stars in the tail of the SMC (also called the wing) have little distance variation, indicating
that it does not have a substantial depth.  On the other hand, at the
tip of the SMC tail/wing, there are two Bridge associations within $17^{\prime}$ (300 pc at 
55 kpc) which are $\approx 5$ kpc 
apart along the line of sight.  In general, the stars in the Bridge show a distance gradient 
expected for a feature
linking the LMC (at 50 kpc) and the SMC (at 60 kpc).  
The stars do not form a continuous
link as the HI does, but are found in loose associations scattered throughout the SMC tail and
decreasing in number towards the central region of the Bridge.

Chemical abundances for the stars in the Bridge were thought to be consistent with an SMC origin
(Rolleston~\etal 1993; Hambly~\etal 1994);
however, recent determinations by Rolleston \& McKenna (1999) suggest they
are deficient by $\sim0.6$ dex compared to similar B-type stars in the SMC.
The ages of the Bridge stars range from
10 - 25 Myr, much younger than expected if they were torn from the SMC 200 Myr ago as most
tidal models predict.  This indicates that the Bridge is actually a star forming
region, but searches for ongoing star formation have not yet been successful.
By considering all of the stars in the Bridge, Grondin~\etal (1992) find that the
Bridge's IMF is shallower than that of the Milky Way or the Clouds.  This favors
the formation of massive stars and may indicate that cloud-cloud collisions
are the dominant star formation trigger (Scoville~\etal 1986; Christodoulou~\etal 1997).  
 There has been no detection of a horizontal branch star population in the Bridge, 
indicating that the halos of 
the two clouds do not meet (Grondin~\etal 1992).  
Kunkel~\etal (1997) have found an abundance of intermediate-age (several Gyr) carbon stars 
scattered throughout the Bridge region, with possible extensions into the beginning of the Stream.
Diffuse $H_{\alpha}$ emission also appears to be prevalent in the Bridge
region closest to the SMC (Johnson~\etal 1982; Marcelin~\etal 1985), as would 
be expected with the presence of hot young stars.  However, there are also several non-detections in
the central region of the Magellanic Bridge (Veilleux et al. 2001).

\begin{figure}
\vspace{5cm}
\caption{Neutral hydrogen column density map of the Magellanic Clouds and Bridge with the main
features discussed in the text labelled.  The intensity scale is logarithmic 
ranging from $10^{21}$ cm$^{-2}$ (black) to $2 \times 10^{18}$ cm$^{-2}$.}

\end{figure}

\section{The Magellanic Stream}

\subsection{HI structure and kinematics}

The Magellanic Stream, discovered 25 yrs ago (Wannier \& Wrixon 1972; Mathewson~\etal 1974), is a complex
arc of neutral hydrogen which starts from the Magellanic Clouds and
trails for over 100$^{\circ}$.  The Stream contains $\approx 2 \times 10^8$ \Msun~ of neutral hydrogen (at an average distance of 55 kpc) and has a velocity
gradient of over 700 km s$^{-1}$ from head to tip, 390 km s$^{-1}$ greater
than that due to Galactic rotation alone.
Recent HIPASS observations of the Magellanic Stream provide almost
a two-fold improvement in spatial resolution over previous survey data, and depict
increasing complexity in the Stream's structure (see Figure 2).  In particular the maps
reveal multiple filaments at the Stream's head, a twisting
ladder structure along the Stream's length, and small dense clouds which extend 20\deg\ from the Stream's main filament.
A broad overview of the HI properties of the Magellanic Stream is presented below.  See Putman et al. (2001) for a full description.   

The beginning of the Stream is rather chaotic, as it spews out from several 
locations north of the SMC and Bridge at $v_{LSR}= 90-240$ \kms~ (see Figures 2 \& 3).  
There is a slight discontinuity in velocity as the HI enters the 
Stream from the Bridge.  Figure 3 shows how the Stream becomes 
more negative in velocity as it extends away from the Clouds, and how there are multiple initial 
filaments which come to a clumpy end at $l \approx -60$\deg\
and $v_{LSR} \approx 85$ \kms.  
  The main filament of the Stream continues towards the South Galactic Pole, where it reaches
0 \kms.\footnote{ 
When the Stream's velocity coincides with that of the Milky Way ($\approx$ 0 \kms),  detailed
information is lost in the Galactic emission and in the data reduction, which has problems when
the emission completely fills the scan (see Putman et al. 2001).}
It then proceeds north to $(\ell,b) \approx$ (90\deg, $-40$\deg),
$v_{LSR} \approx -450$ \kms, and column densities of only a few $\times 
10^{18}$ cm$^{-2}$ (versus a few $\times 10^{19}$ cm$^{-2}$ at the Stream's head).
Relative to the Galactic Center, the
radial velocity of the Stream gradually becomes more negative from the head ($\sim$ 
50 \kms) to the tip ($\sim$ -200 \kms).  

The main filament of the Stream is not as complex as the head, but it is also a 
complicated structure which appears to be made up
of two distinct components.  The splitting of the Stream into two filaments is evident throughout,
but is most obvious beyond the multiple filaments at the Stream's head.  The
two filaments run parallel for the length of the Stream and 
begin to merge towards the tail (much as if one were looking down a long
straight road).  There are also several horse-shoe shaped structures which 
join the two filaments at several positions.  This helical structure
may represent the orbit of the Magellanic Clouds about each
other, with the two filaments representing material from the Bridge and SMC.

Small compact clouds are found throughout Figs. 2 and 3, surrounding the
Stream's main filament in both position and velocity. 
Many of the small clouds, both in and about the Stream, show head-tail structures (i.e. a dense core with 
a diffuse extension of approximately twice the diameter of the core (tadpoles)) and hollow bow-shock
signatures (also noted by Mathewson~\etal (1979)).  This is especially true  
at the Stream's head, with the tails generally pointing away from the Clouds.
This could be depicting the Stream's interaction with the Galaxy's
halo (Pietz~\etal 1996), or it could simply represent the way the gas has been 
stripped from the Clouds.  Some of the small clouds of positive velocity HI about the 
South Galactic Pole in Figs. 2 and 3  
are actually galaxies of the Sculptor Group.  It has been argued that
the abundance of small clouds between these galaxies are not
associated with the Stream, but are
 members of the Sculptor Group (Mathewson~\etal 1975; Haynes \& Roberts 1979).
Considering the Stream's clumpy nature throughout this area, it would be
difficult to make a confident claim of a cloud's association with the Sculptor Group or
other dwarf galaxies (e.g. Carignan~\etal 1998).   However, it is curious how the 
clumps remain in the southern region of the Sculptor Group from velocities 
of $-240$ to $+240$ \kms, and do not follow the Stream as closely in velocity as other clumps
along its length. 
Could these clumps be the remnants of an ejection from the Galactic Centre, or possibly intergalactic
HI clouds along the Coma-Sculptor-Local Group supergalactic filament (Tully \& Fisher 1987; Jerjen, Freeman \& Binggeli 1998)?
H$_\alpha$ observations, metallicity and distance determinations should help distinguish
between these possibilities.

The small-scale HI spatial structure of the Stream has not yet been investigated, but ATCA observations are 
being actively pursued.  HIPASS has a velocity resolution of only 26~\kms~ 
with Hanning smoothing, but higher velocity resolution observations (1~\kms) are in progress with the Parkes 
narrow-band facility (Br\"{u}ens \etal 2000).  Higher velocity resolution observations have also been 
completed in the past by 
Haynes (1979), who noted the complex, multi-profile nature of the Stream
in the region near the South Galactic Pole, by Cohen (1982), who found the Stream to also
have a strong transverse velocity gradient, and by Morras (1983; 1985), who
noted the bifurcation of the Stream.   The northern tip of the Stream was
studied by Wayte (1989).  He notes the continued bifurcation of the Stream
and a complex velocity structure which may indicate
that the tail of the Stream is breaking up into many individual clouds at different 
velocities.  The line profiles of the clouds at the tip show a core/envelope 
structure reminiscent of some non-Magellanic HVCs (Wakker \& vanWoerden 1997).  
If these non-Magellanic HVCs are generally less distant than the majority of
the Stream (see van Woerden \etal (1999) for some distances), this 
change in profile may indicate that the tip of the Stream is getting closer to the Galaxy.
Tidal or ram pressure forces may be responsible for stripping off
the clouds' outer layers.  

\subsection{Optical Observations}

A new method of studying the Magellanic Stream has come with
the discovery that the Stream can be detected in H$\alpha$ emission (e.g. Weiner
\& Williams 1996).  The detections vary tremendously in surface brightness ($0.04 - 0.4$ Rayleighs), and are usually
in regions of high HI column density ($> 10^{19}$ cm$^{-2}$).  There does not appear to be a
correlation between the H$\alpha$ emission measure and HI column density, however
the current lack of high-resolution HI data makes this difficult to test.  
It is possible that H$\alpha$ emission will be detected beyond the HI contours
of the Stream as this has been observed for other complexes (Tufte \etal 1998) and
may indicate the presence of an ionized sheath.
Other lines have also now been detected, including
[NII], [SII], and a non-detection of [OIII] (Bland-Hawthorn~\etal 1999).  
It is not clear how the emission line results should be interpreted.  
Earlier suggestions that ram-pressure is responsible for the H$\alpha$
emission seem less secure in light of the line ratios and the higher
resolution HI maps which show that the strong detections do not always
correlate with the leading edges of HI condensations (Putman \& Gibson 1999).
Bland-Hawthorn \& Maloney (1999) conclude that shock ionization
requires unrealistically high halo densities at $d\approx 50$\,kpc and suggest ionizing photons
from the Galaxy are the main cause for the emission.
On the other hand, the H$\alpha$
emission measures in the Stream are generally $\sim2$ times higher than
 HVCs which have upper distance
limits of $\sim$10\,kpc (Tufte~\etal~1998).  It remains to be seen if 
the contribution of ionizing photons from the LMC, or the effects of shadowing and
nearby spiral arms, can account for this difference.
If the escape of ionizing photons from the Galaxy and the Magellanic Clouds can be 
accurately determined, the emission measures can be used to determine the distance
to various points along the Stream (Bland-Hawthorn \& Maloney 1999). 
A complete map of the Stream's ionized gas would be a very interesting complement to the HI data
presented here.

There have been numerous searches for stars which are associated with the Stream, as they might
be expected if the Stream
were formed via a gravitational interaction.  All of the searches for stars within the
HI contours of the Stream have been negative, with most of the searches being based on
the assumption that the Stream is young and should be populated by A - F stars.  
Br\"uck \& Hawkins (1983) claimed no stellar Stream counterpart based on
star counts down to magnitude 20.5 in B in the
section of the Stream closest to the Clouds.  Recillas-Cruz (1982) and 
Tanaka \& Hamajima (1982) did a similar search of the tip of the Stream and found no excess 
of A-type stars.  
Guhathakurta \& Reitzel (1998) recently used the Keck telescope to complete a deep stellar search in a 
$5^{\prime} \times 7^{\prime}$
region at $(\ell,b) \approx$ 60\deg, -68\deg (within MS IV) and claimed an upper 
limit on the Stream's star-to-gas ratio of 0.1 (5\% that of the LMC). 
It is possible that these results are still not definitive, given the young population of stars searched for 
in the early searches and the limited area covered by the Keck search; but a more likely explanation
 is that the HI Stream does not contain stars.  There is still the
possibility of an offset stellar stream (as seen in many other interacting systems; Hibbard
\& Yun 1999) or a stellar stream that is significantly less extended than the Stream due to the initial HI distribution
of the Clouds being more extended (Yoshizawa 1998).
A possible offset stellar tidal counterpart has been found by Majewski~\etal (1999), who searched for
giant stars about the Clouds and found interesting populations in a region north of the LMC.

\subsection{Metallicity \& Distance Determinations}

Metallicity and abundance determinations for the Magellanic Stream are consistent with a Magellanic
Clouds origin.
The Stream's primary metallicity determination uses Fairall 9 as a probe and has
been investigated by Gibson \etal (2000), Lu \etal (1994) and Songaila (1981), all of whom
obtained consistent results.  Recently, Gibson \etal
used GHRS data and new HI observations to obtain a S/H of 0.21 solar, extremely close to the metallicity
of the SMC.  They also detected Mg II near the tip of the Stream which indicates
that the Stream gas extends at least 15\deg\ from the HI shown in Figure 2.
Lu \etal (1994) found Si/H \simgt\ 0.2 solar 
and S/H \simlt\ 0.3 solar along the Fairall 9 sightline.
They find the subsequent Si/S ratio to be 
 greater than or equal to 0.6 the solar ratio, which indicates that dust depletion 
is not prevalent in the Stream (Si is easily depleted onto dust grains)\footnote{They assumed the
intrinsic Si/S ratio was the same as the Sun's.}.  This is consistent with the lack of
extinction and infrared emission from the Stream (Fong \etal 1987).  The extinction result
is based primarily on galaxy counts (see also Mathewson \etal 1979) and, though inconclusive, 
the results suggest at most a very small level of extinction. 

Sembach \etal (2000) have detected O VI associated with the Magellanic Stream, indicating
that hot gas must be present.  It is very difficult to produce O VI with
photoionization and they suggest movement through a hot Galactic halo
medium may be responsible.
Lu \etal also have a possible detection of C IV absorption at the position 
of Fairall 9.  This suggests, along with the H$\alpha$ detections discussed above,  
that the metallicity estimates are subject to an ionization correction.
Another uncertainty in the metallicity determinations is the HI column 
density.  The above determinations are based on fairly low spatial 
resolution HI data ($15.^{\prime}5$ or $34^{\prime}$), and HVCs have been known to 
vary by a factor of five in column density on scales of only $1^{\prime}$ (WvW97). 
The metallicity determinations remain clear in their indication of the Stream being made up of
non-primordial gas and are consistent with the Stream originating from the Magellanic Clouds.

Distance estimates for the Stream are based largely on
theoretical interaction models (see section 5.2 for a full description).  
Watanabe (1981) made the assumption that the shape of
the Stream clouds (i.e. elongation) is determined by the strength of the Galactic tidal disruption
force and estimates the Stream lies between 36 - 50 kpc. 
H$\alpha$ observations also have the potential to provide distance information (see section 3.2). 

\begin{figure}
\vspace{5cm}
\caption{An integrated intensity map of the Magellanic Stream 
($v_{LSR} = -400$ to $+400$ \kms), which includes the region
shown in Figure 1, part of the Leading Arm shown in Figure 4 and the
full extent of the Magellanic Stream.
The Stream passes through the velocity of Galactic emission at
$(\ell,b) \approx$ 315\deg, -80\deg, and the emission
between +/- 20 \kms in this region has been excluded (see Putman \etal (2000)
for the channel maps).
The intensity values are on a logarithmic scale, with everything above
$6 \times 10^{20}$ cm$^{-2}$ black and the faintest levels at 
approximately $2 \times 10^{18}$ cm$^{-2}$.}
\end{figure}

\begin{figure}
\vspace{5cm}
\caption{Velocity distribution of the Magellanic Stream ranging from
-450 \kms (light grey) to 380 \kms (dark).}
\end{figure}

\section{The Leading Arm}

\subsection{HI structure and kinematics}

The Leading Arm is made up of a string of clouds on the
leading side of the Magellanic Clouds which have only recently 
been clarified as being connected to each other and the Magellanic System
through HIPASS observations (Putman~\etal 1998).  The beginning of the Leading
Arm protrudes from the Magellanic 
Bridge and LMC along several clumpy filaments (see Figure 4).  
The multiple filaments give the appearance that the Leading Arm is associated with both of the Clouds.
The Leading Arm is relatively thin ($\sim 1/4$ the width of the trailing Stream), but roughly 
continuous until the Galactic Plane, where it abruptly shifts in Galactic Longitude from
307$^{\circ}$ to 290$^{\circ}$.  
The Leading Arm is very clumpy, with diffuse
filaments connecting the clumps.  These filaments were missed in previous surveys due to sparse 
spatial sampling (Mathewson \& Ford 1984; Morras 1982), and it was thought
that the clumps were isolated high-velocity clouds.  
There are also dense clouds about the main filament of the Leading Arm (primarily on the
lower longitude side), similar to the small clouds which surround the Stream.  

The Leading Arm's velocity distribution is somewhat confusing, and
this may be due to the projection of the feature.  It emanates
from the Clouds at $v_{\rm lsr}\approx 180$~\kms~and its velocity steadily increases until it reaches 
356 km s$^{-1}$ at $(\ell,b)=(302^\circ,-17^\circ)$.
From this position it decreases in velocity to $\approx 200$~\kms~as it moves towards the 
Galactic Plane (see Putman~\etal (1998) for channel maps).
When the Arm shifts in position by 15$^{\circ}$ in longitude at the Plane, it also shifts in 
velocity, starting at $\approx$ 320~\kms~ at latitude $+8^{\circ}$ and extending to 150~\kms~ at 
latitude $+30^{\circ}$.   Relative to the Galactic Center, the Leading
Arm extends in velocity from $v_{\rm gsr} = -29$ to 178 \kms. 
The metallicity determination discussed below suggests
that the feature at positive latitudes is a continuation of Magellanic material;
however, it is difficult to reproduce the Leading Arm's initial $\sim 60^\circ$ deflection
angle from the great circle defined by the Stream, while also retaining the positive latitude
clouds as tidal debris (Gardiner 1999).
 Verschuur (1975) suggested that the high positive velocity features which make up the Leading
Arm are actually distant spiral features which form an intergalactic bridge between the 
Clouds and the Milky Way.  This seems unlikely since the HIPASS observations show the Leading
Arm's velocity to be
distinct from the velocity of the Galactic HI in this direction ($\sim 120$ \kms; Burton 1988). 

It appears as if the data shown in Figure 4 represent the full extent of the Leading Arm feature, as
maps further north of $b=30^{\circ}$ do not show any obvious continuation of emission.  
It is curious that the
filament abruptly ends at a relatively high column density; however, there could be
more tenuous or fully ionized gas further along. The Leading Arm is not as ordered
or massive as the Stream, possibly due to its leading position or age. 
The mass of the Leading Arm is approximately $2 \times 10^7 M_\odot$, an order of magnitude 
less massive than the Stream, assuming
they are both at the distance of the Magellanic Clouds.

High resolution ATCA observations are in progress for many positions along the Leading Arm.
Wakker~\etal (1999) have already analyzed ATCA data for a position on the positive latitude side 
of the Plane and found velocity widths of 5-10~\kms~ and column density
contrasts of a factor of 3 on arc minute scales.
They also note the two-component velocity structure of this cloud, similar to other non-Magellanic
HVCs, and derive a pressure of 18000 R$^{-1} D^{-1}_{kpc}$ K cm$^{-3}$ (where R is the resolution in arc minutes and D is the distance to the cloud).  
Other observations of the Leading Arm include the work of Bajaja~\etal (1989), Morras \& Bajaja 
(1983) and Morras (1982); all 
of which are at a lower spatial resolution but higher velocity resolution than the data shown here.

\subsection{Metallicity Determination}

Apart from the fact that the Leading Arm emanates from the Magellanic System, the
strongest evidence that it is made of Magellanic material comes from the Lu~\etal (1998) metallicity
determination for HVC 287.5+22.5+240.  
Derived from GHRS spectra of the background galaxy NGC 3783 (see Figure 4 for position), a S/H of $\approx 0.25\odot$
 was found, consistent with the metallicity of the Magellanic Clouds.
They also found Fe/H $= 0.033\odot$, with the subsolar Fe/S ratio indicating
dust may be present.
This filament lies spatially (and kinematically) in a region where tidal models predict
 gaseous tidal debris to reside, and the metallicity determination suggests that despite the offset
positioning of this filament, it is indeed part of the Magellanic Leading Arm.
The position of the Seyfert galaxy ESO265-G23 is another possible background source
which can be used to determine the Leading Arm's metallicity; however it
appears to be just off the HI contours in the HIPASS map (see Putman \& Gibson
1999).  This position may either have a very low column density or represent the ionized
medium of the Arm.

\begin{figure}
\vspace{5cm}
\caption{A HIPASS peak intensity map which shows the full extent of the Leading Arm, as well as 
the Magellanic Clouds, the Bridge and the beginning of the Stream (as labelled).  The position
of the background galaxy, NGC 3783, is also noted (see $\S$4.2).  To avoid the emission from the Galactic Plane (which extends out
to 120 \kms in this direction), only velocities between 130 and 400 \kms were used.  (Thus
the strange appearance of the SMC which begins at $\approx 80$ \kms.)  Many 
features are intentionally saturated to bring out the low level emission.  It is
a linear intensity scale ranging from approximately 0.1 to 2 K (black).}
\end{figure}

\section{Theoretical Origin Models}

\subsection{Bridge}
It is generally agreed that the LMC and SMC are bound and that the Bridge was
formed via a tidal encounter between the two Clouds (e.g. Gardiner \& Noguchi 1996 
(hereafter GN96); Moore \& Davis 1994).  
The finding of stars in the Bridge region supports the tidal model, though the young
stellar population may have been born in the Bridge (see $\S$2.2).  
Few models can simultaneously reproduce both the Bridge and the Stream accurately.
GN96 are relatively successful by refining the models of Lin \& 
Lynden Bell (1982) and Murai \&
Fujimoto (1980).  They find that the Magellanic Bridge was most likely pulled from
the SMC 0.2 Gyr ago during a close encounter between the two Clouds (at 7 kpc separation). 
The GN96 model, in which the SMC is composed of both a disk and
a halo, nicely explains the different bridge and tail HI components and the velocity 
distribution of the young (early-type) and old (carbon star) stellar populations.

In contrast to GN96, Kunkel~\etal (1994) attempt to reproduce the properties of the SMC and Bridge
by leaving the LMC
and SMC unbound and ignoring the effect of the Galaxy (i.e. they do not reproduce
the Magellanic Stream).  They suggest that the carbon stars are part of the tidal bridge,
separate from the HI and embedded in some type of ionized medium.
Heller \& Rohlfs (1994) agree with GN96 that the two Clouds are bound, but
argue that they have
remained in a stable binary system for the last $10^{10}$ years and that tidal forces from the Galaxy 
were not strong enough to pull out the Magellanic Stream.
They suggest the Bridge or intercloud region was formed 0.5 Gyr ago when there was a close
encounter between the LMC and SMC, and this also
marks the beginning of the formation of the Magellanic Stream.
The chaotic nature of the HI features north of the Bridge (at the head of the Stream) indicates that
the Bridge and Stream were not formed in conjunction or that one is pulling material from the other.
In the best model of Li (1999), the Clouds have only been gravitationally
affecting each other for the past 2 Gyr, as he also finds that when the
Clouds are a lifelong binary, the interaction between the two Clouds does
not allow the Magellanic Stream to form.

All of the models assume that the Bridge is made up of material from the SMC, with the 
LMC ripping material from its less
massive companion.  The HIPASS data show an extension of the LMC which suggests that the 
LMC also contributes to the Bridge's mass (see Figure 1).  
This feature may be reproduced when the potential of the LMC is modelled more realistically.
 
\subsection{Stream and Leading Arm}

The Stream is the result of an interaction between the Galaxy and the Clouds; its
link to the Magellanic System and spatial and kinematic continuity
are the primary clues for this conclusion.  The exact
form of the interaction is not yet fully understood, but its striking
appearance has attracted an abundance of theoretical attention.
Many of the early models were created before the tangential velocity of 
the Clouds was known\footnote{Proper motion measurements have now shown the  
Clouds are leading the Stream with a total galactocentric transverse velocity of 215$\pm48$ \kms (Jones et al.
1994).}, or they were simply unable to reproduce the observed data.
This section summarizes
the more recent developments in our theoretical understanding of the Magellanic HVC's
formation and evolution.
The models have generally been variations on two themes:  gravitational tides from the
Milky Way pulling 
the Stream from the Clouds, and ram-pressure stripping of the Stream gas as the Clouds interact with
some form of Galactic gas.  
The finding of the tidal Leading Arm feature (Putman~\etal 1998) indicates that
tidal forces are the dominant mechanism responsible for the formation of the Stream, 
but it is likely that other mechanisms also play a role in the Stream's evolution.

\subsubsection{Tidal Models}
Tidal models have gradually become more complex to match the increasing detail
revealed in the observations.
One of the most recent and advanced
N-body tidal models is that of GN96 which simulates
the SMC as a collection of self-gravitating particles and the LMC as a point mass.   
GN96 is an adaptation of early tidal models (e.g. Murai \& Fujimoto 1980;
Lin \& Lynden-Bell 1982), where the Clouds are in a polar orbit leading
the Stream and are presently close to perigalacticon (see also Gardiner~\etal 1994; Lin~\etal
1995).  To achieve the high negative velocities
at the Stream's tip, these models invoked a Galaxy with a massive halo ($\sim 10^{12} M_{\odot}$) 
which extends out to $\sim 200$ kpc, consistent with recent results (e.g. Kochanek 1996).
GN96 (and other recent tidal models) predict that the Stream was
pulled from the SMC 1.5-2 Gyr ago, when a tidal encounter between the two Clouds (at 14 
kpc separation) coincided with their previous perigalactic passage.
The Stream was drawn into its present position as the Clouds moved from 
apogalacticon ($\sim0.9$ Gyr ago) to their present position, just past perigalacticon.
GN96 find that the Stream consists of two separate streams, a main filament along the
observed position of the MS and a less densely populated secondary filament (this
secondary stream is also a prediction of Tanaka (1981)).  This splitting of the
Stream is seen in the HIPASS data shown in Figs. 2 and 3; however the separation of the
two components is significantly larger in the model and the cause of the separation
remains unclear.  The dual filaments may represent multiple close encounters between the two Clouds, resulting in 
two major gas concentrations which were subsequently drawn into the Stream.
The fact that the filaments are at approximately the same velocity argues for a similar origin.
GN96 reproduce the velocity distribution of the Stream fairly accurately (see Figure 5), but the
variation in column density along the Stream requires further work. 

\begin{figure}
\vspace{5cm}
\caption{The velocity profile of the particles in the best
tidal plus weak drag model of Gardiner (1999)
with the observational data of Mathewson \etal (1974) included for comparison.  
The velocities are shown in the galactic standard of rest (GSR) reference frame, and
the coordinates are Magellanic longitude as defined by Wannier \& Wrixon (1972).
The Stream extends from Magellanic longitude $-25$\deg~to 100\deg~in this
coordinate system.}
\end{figure}

An advancement on GN96 has been developed by Yoshizawa (1998), who 
incorporates gas dynamics (via a sticky-particle method) and star formation into the numerical code
(see Figure 6).  The simulations find the beginning of the Stream to consist of multiple filaments, much as depicted
in Figs. 2 and 3.  The simulated Stream then becomes very narrow due to gas dissipation from
cloud-cloud collisions, and the bifurcation found in previous models is lost.  
An important result of the Yoshizawa models is the demonstration
that stars should \it not \rm be drawn out along the Stream,
but remain restricted to a
$\sim 10-15$ degree region surrounding the Clouds (appearing clump-like, or
perhaps in several dispersed streams).  The lack of stars in the Stream
has been a major argument against the Stream having a tidal origin (e.g. Moore \& Davis 1994).
Yoshizawa (1998) preferentially disrupts the gas 
 by having the initial gas distribution of the Clouds
more extended than the stellar component (a common occurrence - e.g.
Broeils \& van Woerden 1994; Yun~\etal 1994).  
As mentioned in $\S$3.2, recent observational work indicates that there may be an excess of 
giants at distances
expected for tidal debris from the Magellanic Clouds and distributed in patterns suggestive
of the small stellar streams predicted by Yoshizawa's models (Majewski~\etal 1999).

\begin{figure}
\vspace{5cm}
\caption{Yoshizawa's (1998) tidal simulation aimed at reproducing the Magellanic HVCs.  The panel
on the left shows the distribution of gas particles, while the right shows the distribution
of star particles.  The SMC is at $(\ell,b) \approx$ 303\deg, -44\deg and the LMC is
at $(\ell,b) \approx$ 280\deg, -33\deg (marked with the open circle).}
\end{figure}
 
A natural result of the tidal model is a leading counterpart to the Stream, the Leading Arm.
The original tidal models presumed that the interaction could be represented
as a two-body problem between the Galaxy and the LMC, which resulted in symmetric
leading and trailing streams of material.  The more recent models of GN96 and
Yoshizawa (1998) treat the interaction as a more realistic 3-body problem (Galaxy, LMC
and SMC) and the perturbative nature of the LMC+SMC interaction leads to HI features
which are clearly non-symmetric.  The strong gravitational perturbation of the LMC 
pulls most of the material in the leading section back towards the Clouds, leaving
a much weakened leading feature compared to the Stream.
GN96 predicts a leading arm which, between the
Magellanic Clouds and the Galactic Plane, has a mass $\sim 1/3$ that of
the entire trailing Stream, a relatively flat velocity gradient, and a
deviation from the Great Circle defined by the trailing Stream of
$\sim30^\circ$.  The newly-discovered Leading Arm has a
mass $\sim 1/10$ that of the Stream (assuming the Arm and the Stream are at the same
distance) and a deviation angle closer to $\sim 60^\circ$.
Though the Leading Arm does not match the predictions of the tidal models
exactly, there are several additions to the current models which could
change this situation.
The differences in the mass and projected orientation could be due to
the shape of the LMC's potential in tidal models (presently a rigid spheroid),
a triaxial distribution of Galactic halo mass (Lin~\etal 1995), and/or a perturbation by another
satellite of the Milky Way (e.g. the Sgr dwarf).  The addition
of a small amount of drag to the tidal model (Gardiner 1999) is able to
reproduce the angle of deflection from the Stream's Great Circle and
the velocity distribution of the Leading Arm (see Figure 5), but it also introduces
an extended anomalous component which wraps around to join the Stream and
is not observed.  
The hydrodynamical models of Li (1999) indicate
that a tidal interaction is not a tidy process and that multiple clumps
of material would be drawn from the Clouds, along with the continuous streams.  This 
could explain the rest of the debris seen in Figs. 2 and 3.  They also find
that the LMC has a substantial effect on the distribution of the leading gas, and that
the stellar component of the Clouds remains largely confined.


\subsubsection{Ram-Pressure Models}
As noted above, combining aspects of the ram-pressure models to the
tidal ones may be the key to reproducing all of the observational
features of the Stream and Arm.   
Mathewson, Schwarz \& Murray (1977) were the first to suggest that
the Stream was formed via thermal instabilities in the wake of the Clouds
during their passage through the Galaxy's hot halo.  These instabilities
form cold clouds which lose their buoyancy and sink towards the Galactic
center.  Variations on this model were subsequently developed and 
simulated.  Liu (1992) proposed cold gas from the Clouds 
was dragged into their wake, and gravitational forces
from the Milky Way accelerated the gas down the vortex. 
Meurer \etal (1985) simulated the tearing of cloudlets from the Bridge as the
Magellanic Clouds passed through the hot gaseous halo of our Galaxy and 
they stretched out the Stream with tidal and drag
forces.  Meurer \etal are able to produce a reasonable model of the
Stream (magnitude of the spatial and velocity extent within a factor
of 2) with a broad range of parameters, but their best model puts
the Stream clouds at an average distance of 38 kpc.  
Sofue (1994) also produces the Stream by passing the Clouds through 
Galactic halo and disk gases and elongates it with the Galaxy's potential.  A
leading stream is formed in the Sofue model when the Stream
begins to rotate around the Galaxy at a higher angular velocity than
the LMC and it wraps all the way around; but this
is at the expense of the predicted extent and velocity
profile of the Stream.  In all of Sofue's simulations the Stream accretes
onto the Galaxy within a few Gyr.  

Moore \& Davis (1994) have a similar, but more detailed model compared to the Meurer \etal and Sofue models.
They pass the Clouds through an extended ionized Galactic disk which
strips off $20\%$ of the Clouds' least bound HI into the Stream.  The
main interaction is thought to have taken place 0.5 Gyr ago at a
distance of 65 kpc and they propose that the material responsible for
the stripping is an extension of the Galactic  HI disk which has
column densities $< 10^{19}$ cm$^{-2}$ and is ionized by the
extragalactic background radiation.
Moore \& Davis are able to explain the Stream's column density
gradient and the high negative velocities at the Stream's tip, as the
gas with the lowest column density loses the  most orbital angular
momentum and falls to a distance of ~20 kpc from the Galaxy with a
velocity  of -380 \kms~ ($v_{\rm lsr}$).  Without the addition of a braking effect from an
extended dilute halo of ionized gas, the stripped material actually
begins to lead the Magellanic Clouds.
This is because the gas clouds lose energy when initially stripped from
the Clouds, which causes the apogalactic distance of their orbit to decrease.
As their orbital period decreases they overtake the Clouds as projected
on the sky.

A slightly different approach to the ram pressure model was taken by 
Heller \& Rohlfs (1994) and Mathewson \etal~(1987).  Heller \& Rohlfs mark
the beginning of the formation of the Stream at 0.5 Gyr ago when the
Magellanic Clouds had a close encounter with each other and much of the HI was
disrupted from the core of the LMC and SMC into the Bridge region or
some other extended configuration.  The Stream was subsequently swept
out by a strong wind generated by the orbital motion of the Clouds
through the Galactic halo.  Mathewson \etal\ (1987) propose a discrete 
ram-pressure model in which
the Stream's formation is due to the Clouds' interaction with
high-velocity clouds presently found on the leading side of the
Clouds.  Detailed results are not available, but the fact that these
leading clouds are shown here as the continuous Leading Arm
casts doubt on this model.

\bigskip
Besides the further development and combination of the models currently
in existence, there are several observational tests which can be carried
out to distinguish between the origin scenarios.  As previously discussed, 
continuing the search for a stellar counterpart to the HI Stream and Leading Arm
is of importance.  Abundance determinations will also
be a crucial tool for confirming the origin of some of the more remote
clouds which are proposed members of the Magellanic System.  Searching for
soft x-ray emission along the Stream and investigating the optical line ratios may provide insight
into the mechanisms responsible for putting the Stream in its current state.
Understanding the ionized component of the Stream is an important future goal.
If it is determined that photoionization is the dominant ionizing
mechanism, mapping the Stream in H$\alpha$ emission has the potential to
reveal the three-dimensional distribution of the Stream (Bland-Hawthorn \& 
Maloney 1999; Bland-Hawthorn~\etal 1999; see also $\S$3.2).  Because ram
pressure models put the tip of the Stream at $\sim 20 - 50$\,kpc (e.g. Moore \&
Davis 1994;  Heller \& Rohlfs 1994), and tidal models put the tip at $\sim 70 - 100$\,kpc (e.g. Gardiner
1999), H$\alpha$ observations of MS VI, in particular, might provide an elegant
test for the competing models.
It will be difficult to reproduce the detail revealed in the HIPASS Stream observations,
but the gross properties should be matched before any interaction model is adopted. 
These properties include:  a chaotic beginning consisting of multiple filaments,
dual streams, narrowing of the main filament towards the tip and a broadening
at the very end, compact clouds which surround the main filament, and a
continuous velocity structure.  

\section{Relationships to other HVCs?}

It has been suggested that the complexes described here are what is left of a polar ring 
of Magellanic debris (e.g. Mathewson~\etal  1987).
In fact, it is still possible that all HVCs originated as part of the Magellanic Cloud/Milky
Way interaction (Mathewson~\etal 1974), but the evidence points against it considering the survival
timescales, HVC abundance and distance measurements (Wakker \& van Woerden 1997), and the 
predicted position and velocity
distribution of the Magellanic remnants (Wakker \& Bregman 1990).  
A more likely possibility is that other high velocity complexes are the remnants
of previous non-Magellanic interactions and/or torn apart Galactic satellites.  
When developing models to explain the global HVC population, it is 
important that the HVCs which are known to be part of the Magellanic System are excluded.  For instance the identification of the Leading
Arm eliminates many of the extreme positive velocity clouds, and the dense clumps
about the Magellanic Stream are also likely to be interaction related and 
should be excluded from analyses which attempt to match such clouds to a Local
Group origin (e.g. Braun \& Burton 1999).
When the Magellanic HVCs are no longer included in the overall population of HVCs, the 
sky covering fraction goes down by at least 5\% (Wakker \& van Woerden 1997).  It also leads
to a serious deficiency of HVCs in the southern sky and at positive velocities.
If most of the positive velocity HVCs can be classified as Magellanic debris or
Galactic extensions we may be able to reconsider various origin scenarios which
are unable to produce high positive velocity gas. 
 
In the quest to understand the global origin of HVCs, one of the most important roles of 
the Magellanic complexes is to provide a basis for observational comparisons.
The major HVC questions concern their origin and environment, and 
 are highly dependent on the clouds' distances.  Since the Magellanic HVCs
are known to have originated from the Clouds and lie at distances in the
tens of kiloparsecs, they can be used as a calibrator to investigate the HVC phenomena.  
To begin, the overall spatial structure of the Magellanic complexes and other HVCs has similarities 
and differences on both large and small scales.
The long filamentary structure of the Stream and Arm appears to be common throughout HVCs and may
indicate that more clouds either have a tidal origin or are currently being 
tidally stretched.  If assumptions are made about the 
density fall-off of the halo, the length of the
head-tail and bow shock structures may allow an estimate of the clouds' distances (Mebold, pers. comm.).  
The small scale structure of HVCs supplies information about the
physical conditions within the clouds, and an indication of the amount of
turbulence or ordered structure.  The future high resolution HI observations of the
Magellanic complexes should be compared to similar HVC observations (e.g. Wakker 
\& Schwarz 1991).
The Magellanic Bridge is useful as it
can be used to examine the effects of star formation on HI structure and to explore
what has triggered this process.  
The kinematic structure of the HI can also provide clues to the formation and evolution
of HVCs.  While many of the Magellanic HVCs show a single component line profile,
other HVCs show two-component profiles, indicative of a core-envelope structure or a 
two-phase medium.
The Magellanic Stream and Bridge also show a systematic variation in velocity and column density
which is uncommon in high velocity complexes.   There are 
useful comparisons to made at all wavelengths, including optical emission
and absorption line observations to determine abundances and ionization conditions.
 These types of comparisons will 
clarify some of the current unknowns about the Magellanic complexes and their relation to 
the entire population of high-velocity clouds.

\section{Acknowledgements}
I would like to thank Ken Freeman, Lister Staveley-Smith, and Brad Gibson for providing useful feedback
on the chapter.  I would also like to thank Lance Gardiner and Akira Yoshizawa and the Astronomical Society of the Pacific 
Conference Series for permission to reproduce Figure 5 and 6.  The Australia
Telescope National Facility is thanked for hosting me during much of this work.

{}


\begin{thebibliography}{} 
\bibitem[]{}
Bajaja, E., Cappa de Nicolau, C.E., Martin, M.C., Morras, R., Olano, C.A. \& Poppel, W.G.L.
1989, A\&A Suppl, 78, 345

\bibitem[]{} Barnes, D.G. et al. 2000, MNRAS, accepted


\bibitem[]{}
Bland-Hawthorn, J. et al. 1999, ApJ, in preparation
\bibitem[]{}
Bland-Hawthorn, J. \& Maloney, P.R. 1999, ApJ, 510, 33
\bibitem[]{}
Bland-Hawthorn, J., Veilleux, S., Cecil, G.N., Putman, M.E., Gibson, B.K. \& Maloney, P.R. 
1998, MNRAS, 299, 611

\bibitem[]{}
Braun, R. \& Burton, W.B. 1999, A\&A, 341, 437

\bibitem[]{}
Broeils, A.H. \& van Woerden, H. 1994, A\&A Suppl., 107, 129

\bibitem[]{}
 Br\"{u}ck, M.T. \& Hawkins, M.R.S. 1983, 
A\&A, 124, 216 

\bibitem[]{}
 Br\"{u}ens C., Kerp J., Staveley-Smith, L. 2000, in ``Mapping the Hidden Universe'',
ASP Conf. Ser., in press

\bibitem[]{}
 Burton, W.B. 1988,
in {\it Galactic and Extragalactic Radio Astronomy}, eds. Verschuur, G.L.
\& Kellermann, K.I., 295, Springer-Verlag

\bibitem[]{}
Carignan, C., Beaulieu, S., Cote, S., Demers, S. \& Mateo, M. 1998 
 ApJ, 116, 1690
\bibitem[]{}
Cohen, R.J. 1982, MNRAS, 199, 281
\bibitem[]{}
Christodoulou, D.M., Tohline, J.E., Keenan, F.P. 1997, ApJ, 486, 810

\bibitem[]{}
Demers, S. \& Battinelli, P. 1998, AJ, 115, 154


\bibitem[]{}
Fong, R.~\etal 1987 
MNRAS, 224, 1059

\bibitem[]{}
Gardiner, L.T. 1999, in {\it Stromlo Workshop on High-Velocity
Clouds}, ASP Conf. Series \#166, 292

\bibitem[]{}
 Gardiner, L.T. \& Noguchi, M. 1996,
MNRAS, 278, 191 

\bibitem[]{}
Gardiner, L.T., Sawa, T. \& Fujimoto, M. 1994,
MNRAS, 266, 567

\bibitem[]{}
Gibson, B.K., Giroux, M.L., Penton, S.V., Putman, M.E., Stocke, J.T. \&
Shull, J.M. 2000, ApJ accepted

\bibitem[]{}
Grondin, L., Demers, S. \& Kunkel, W.E. 1992, AJ, 103, 1234

\bibitem[]{}
 Guhathakurta, P. \& Reitzel, D.B. 1998,
in {\it Galactic Halos:  A UC Santa Cruz Workshop}, ed. Zaritsky, D., ASP Conf Series \#136

\bibitem[]{}
Haffner, L.M., Reynolds, R.J. \& Tufte, S.L. 1999, ApJ, accepted

\bibitem[]{}
Hambly, N.C., Dufton, P.L., Keenan, F.P., Rolleston, W.R.J., Howarth, I.D. \&
Irwin, M.J. 1994, A\&A, 285, 716

\bibitem[]{}
Haynes, M.P. \& Roberts, M.S. 1979, ApJ, 227, 767

\bibitem[]{}
Haynes, M.P. 1979, AJ, 84, 1173

\bibitem[]{}
Heller, P. \& Rohlfs, K. 1994, A\&A, 291, 743

\bibitem[]{}
Hibbard, J.E. \& Yun, M.S. 1999, ApJ submitted

\bibitem[]{}
Irwin, M.J., Demers, S. \& Kunkel, W.E. 1990, AJ, 99, 191 

\bibitem[]{}
Jerjen, H. Freeman, K. \& Binggeli, B. 1998, AJ, 116, 2873

\bibitem[]{}
Johnson, P.G., Meaburn, J. \& Osman, A.M. 1982, MNRAS, 198, 985

\bibitem[]{}
Jones, B.F., Klemola, A.R. \& Lin, D.N.C. 1994, AJ, 107, 1333

\bibitem[]{}
Kim, S., Staveley-Smith, L., Dopita, M.A., Freeman,
K.C., Sault, R.J., Kesteven, M.J. \& McConnell, D. 1999,
ApJ, 503, 729
 
\bibitem[]{}
Kobulnicky, H.A. \& Dickey, J.M. 1999, AJ, 117, 908

\bibitem[]{}
Kochanek, C.S. 1996, ApJ, 457, 228

\bibitem[]{}
Kunkel, W.E., Demers, M.J. \& Irwin, M.J. 1994, in {\it Proceedings from
the CTIO-ESO Workshop on the Local Group in La Serena, Chile}

\bibitem[]{}
Kunkel, W.E., Irwin, M.J. \& Demers, S. 1997, A\&AS, 122, 463

\bibitem[]{}
Li, P.S. 1999, PhD Thesis, University of Wyoming


\bibitem[]{}
 Lin, D.N.C., Jones, B.F. \& Klemola, A.R. 1995,
ApJ, 439, 652

\bibitem[]{}
Lin, D.N.C. \& Lynden-Bell, D. 1982, MNRAS, 198, 707
\bibitem[]{}
Liu, Y. 1992, A\&A, 257, 505

\bibitem[]{}
 Lu, L., Savage, B.D., Sembach, K.R., Wakker, B.P., Sargent, W.L.W. \&
Oosterloo, T.A. 1998,
AJ, 115, 162
 
\bibitem[]{}
Lu, L., Savage, B.~D. \& Sembach, K.~R. 1994, 
ApJ, 437, L119

\bibitem[]{}
Marcelin, M., Boulesteix, J. \& Georgelin, Y. 1985, Nature, 316, 705

\bibitem[]{}
 Mathewson, D.S., Wayte, S.R., Ford, V.L. \& Ruan, K. 1987,
PASA, 7, 19

\bibitem[]{}
Mathewson, D.S., Schwarz,  M.P. \& Murray, J.D. 1977, PASA, 3, 133

\bibitem[]{}
 Mathewson, D.S. \& Ford, V.L. 1984,
in {\it Structure and Evolution of the Magellanic Clouds} eds. van den Bergh, S. \& de
Boer, K.S., 125, Kluwer, Dordrecht

\bibitem[]{}
Mathewson, D.S., Ford, V.L., Schwarz, M.P. \& Murray, J.D. 1979, in {\it The
Large-scale Structure of the Galaxy}, 556, Dordrecht

\bibitem[]{}
Mathewson, D.S., Cleary, M.N. \& Murray, J.D. 1975, ApJ, 195, 97

\bibitem[]{}
Mathewson, D.S., Cleary, M.N. \& Murray, J.D. 1974, 
AJ, 190, 291 

\bibitem[]{}
McGee, R.X. \& Newton, L.M. 1986, 
PASA, 6, 471

\bibitem[]{}
Majewski, S.R., Ostheimer, J.C., Kunkel, W.E., Johnston, K.V. \& Patterson, R.J. 1999, in {\it IAU Symp. 190: New Views
of the Magellanic Clouds}, ASP Conf. Series, in press

\bibitem[]{}
Meurer, G.R., Bicknell, G.V. \& Gingold, R.A. 1985, PASA, 6, 195

\bibitem[]{}
Moore, B. \& Davis, M. 1994,
MNRAS, 270, 209

\bibitem[]{}
Morras, R. 1985, AJ, 90, 180

\bibitem[]{}
Morras, R. 1983, AJ, 88, 62

\bibitem[]{}
Morras, R. \& Bajaja, E. 1983, A\&AS, 51, 131

\bibitem[]{}
Morras, R. 1982, A\&A, 115, 249
\bibitem[]{}
 Murai, T. \& Fujimoto, M. 1980,
PASJ, 32, 581 

\bibitem[]{}
Oort, J. H. 1970, A\&A, 7, 381

\bibitem[]{}
Pietz, J., Kerp, J., Kalberla, P.M.W., Mebold, U., Burton, W.B. \&
Hartmann, D. 1996, A\&A, 308, 37

\bibitem[]{}
Putman~\etal 2001, AJ, in preparation
\bibitem[]{}
Putman, M.E. \& Gibson, B.K. 1999, PASA, 16, 70 

\bibitem[]{}
Putman, M.E. \& Gibson, B.K. 1999, in {\it Stromlo Workshop on High-Velocity
Clouds}, ASP Conf. Series \#166, 276


\bibitem[]{}
Putman, M.E., Gibson, B.K. \& Staveley-Smith, L. 1999, in {\it IAU Symp. 190: New Views
of the Magellanic Clouds}, ASP Conf. Series, in press
\bibitem[]{}
Putman, M.E., Gibson, B.K, Staveley-Smith, L. \etal~ 1998, Nature, 394, 752

\bibitem[]{}
Recillas-Cruz, E. 1982, MNRAS, 201, 473

\bibitem[]{}
Rolleston, W.R.J. \& McKenna, F.C. 1999, in {\it IAU Symp. 190: New Views
of the Magellanic Clouds}, ASP Conf. Series, in press

\bibitem[]{}
Rolleston, W.R.J., Dufton, P.L., Fitzsimmons, A., Howarth, I.D. \& Irwin, M.J.
1993, A\&A, 277, 10

\bibitem[]{}
Sanduleak, N. 1969, AJ, 74, 877

\bibitem[]{}
Sahu, M.S. 1998, 
AJ, 116, 1205

\bibitem[]{}
Scoville, N.Z., Sanders, D.B. \& Clemens, D.P. 1986, ApJ, 310, 77

\bibitem[]{}
Sembach, K.R., Savage, B.D., Lu, L. \& Murphy, E.M. 1999, ApJ, 515, 108

\bibitem[]{}
Sembach, K.R. et al. 2000, ApJ, 538, 31

\bibitem[]{}
 Sofue, Y. 1994,
PASJ, 46, 431 

\bibitem[]{}
Stanimirovic, S., Staveley-Smith, L., Dickey, J.M., Sault, R.J. \& Snowden, S.L. 1999, MNRAS, 302, 417 
 
\bibitem[]{}
Staveley-Smith, L. 1997, 
PASA, 14, 111

\bibitem[]{}
Tanaka, K.I. 1981, PASJ, 33, 247

\bibitem[]{}
Tanaka, K.I. \& Hamajima, K. 1982, PASJ, 34, 417

\bibitem[]{}
Tufte, S.L., Reynolds, R.J. \& Haffner, L.M. 1998, ApJ, 504, 773

\bibitem[]{}
Tully, R.B. \& Fisher, J.R. 1987, Nearby Galaxy Atlas (Cambridge; Cambridge
Univ. Press)


\bibitem[]{}
van Woerden, H., Schwarz, U.J., Peletier, R.F., Wakker, B.P. \& Kalberla, P.M.W. 1999, Nature, 400, 138

\bibitem[]{}
Veilleux, S., Putman, M.E. \& Bland-Hawthorn, J. 2001, in preparation

\bibitem[]{}
Verschuur, G.L. (1975) 
ARA\&A, 13, 257

\bibitem[]{}
Wakker, B.P., Savage, B.D., Oosterloo, T.A. \& Putman, M.E. 1999, in {\it The
Third Stromlo Symposium:  The Galactic Halo}, eds. Gibson, B.K., Axelrod, T.S.
\& Putman, M.E., ASP COnf. Series \# 165, 120
\bibitem[]{}
Wakker, B.P \& van Woerden, H. 1997, ARA\&A, 35, 217 

\bibitem[]{}
Wakker, B.P. \& Shwarz, U. 1991, A\&A, 250, 484

\bibitem[]{}
Wakker, B.P. \& Bregman, J., 1990, in {\it Interstellar Neutral Hydrogen at High Velocities}, PhD Thesis BP Wakker,
Rijks Univ. Groningen, ch. 5

\bibitem[]{}
Wakker, B.P \& van Woerden, H. 1991, A\&A, 250, 509
\bibitem[]{}
Wannier, P., \& Wrixon, G. T. 1972, ApJL, 173, L119

\bibitem[]{}
Watanabe, T. 1981, AJ, 86, 30

\bibitem[]{}
Wayte, S.R. 1989, PASA, 8, 195

\bibitem[]{}
Weiner, B.J. \& Williams, T.B. 1996, AJ, 111, 1156

\bibitem[]{}
Westerlund, B.E. \& Glaspey, J. 1971, A\&A, 10, 1

\bibitem[]{}
Wiseman, J. J. \& Ho, P. T. P. 1998, ApJ, 502, 676

\bibitem[]{}
Yoshizawa, A. 1998, PhD thesis, Tohoku University, Sendai, Japan

\bibitem[]{}
 Yun, M.S., Ho, P.T.P. \& Lo, K.Y. 1994,
Nature, 372, 530






\end{thebibliography}
\end{document}